\documentclass[11pt,a4paper]{article}
\usepackage[margin=30mm]{geometry}

\usepackage{siunitx}
\usepackage{tocbasic}
\usepackage{graphicx}
\usepackage[super]{natbib}
\usepackage{url}
\usepackage{textgreek}

\DeclareUnicodeCharacter{03B2}{\textbeta}
\DeclareUnicodeCharacter{2212}{--}
\DeclareUnicodeCharacter{2009}{\ }

\title{IR Spectroscopic Studies of Gas-Phase Peptides}
\author{Åke Andersson \and Vitali Zhaunerchyk}
\date{May 15, 2024}
\begin{document}
	
	\pagenumbering{gobble} 
	\maketitle
	
	\begin{abstract}
		Proteins are vital biological molecules found in every living organism, and their function is determined by what shape they fold into. Peptides are essentially subsets of proteins, and therefore ideal as model systems for protein folding. The structure of a molecule is closely related to its vibrational absorption spectrum, which lies in the infrared (IR) range. However, \emph{in vivo} IR spectroscopy is hindered by interference from the surrounding water. Therefore, peptides are preferably studied isolated from solution, in the gas phase. This chapter summarizes the recent IR spectroscopy studies of gas-phase peptides. The collected works show that IR spectroscopy combined with quantum chemical calculations is a powerful tool for deducing the molecular structure. Moreover the wealth of experimental spectra makes possible the evaluation of different quantum chemical models, which can be applied to the larger proteins.
	\end{abstract}
	
	\newpage
	{
		\small
		\tableofcontents
	}
	\newpage
	\pagenumbering{arabic} 
	\setcounter{page}{1}
	
	\section{Introduction}
	Proteins are complex biological molecules found in every living organism. They are composed of long chains of amino acids linked together by peptide bonds. There are 22 proteogenic amino acids, each with a unique side chain. The specific sequence of these amino acids, known as the \emph{primary sequence}, determines how the protein folds into its three-dimensional structure. However, the exact mechanism by which the primary sequence dictates this three-dimensional structure is still not fully understood. The only theoretical methods that are cheap enough to be applied to whole proteins are highly parameterized models trained on empirical data, notably AlphaFold. \cite{jumper2021applying,jumper2021highly,abramson2024accurate} This drives a demand for highly accurate measurements of model systems of proteins. With this demand in mind, spectroscopists have turned to peptides, which are short subsets of proteins that are readily amenable to synthesis, and therefore ideal model systems. Although proteins are naturally found in water, peptides are for this purpose investigated in the gas phase. This allows for the study of intrinsic properties, and makes the experiment more readily comparable with quantum chemical calculations.
	
	Since the molecular structure is intimately related to its vibrational spectrum, which infrared (IR) spectroscopy probes, the latter can serve as a tool for molecular structure elucidation. However, biomolecules in the gas phase have a number density too small for direct IR absorption spectroscopy; too few photons are absorbed. Instead, IR \emph{action spectroscopy} is employed, wherein the excited molecules are counted in some fashion. IR action spectroscopy has been applied to a variety of biomolecules in the gas phase, including lone amino acids,\cite{wu2008investigation,spieler2018vibrational} clusters of amino acids,\cite{yin2015structure,feng2016structure,seo2018side,ma2018structural,andersson2020structure} and peptides of various sizes.\cite{seaiby2016ir,deblase2017conformation,batoon2019characterization,andersson2023indication,stroganova2022structural,sherman2022conformational,yatsyna2019conformational,yatsyna2019competition,batoon2018conformations} 
	
	The core of an IR gas-phase spectroscopy study is the interaction between a molecule and some photons. In our case the interaction is absorption, the molecule is a charged or neutral peptide, and at least one photon is IR. The charge state and the combination of photons absorbed define a spectroscopy type. This chapter aims to concisely summarize the recent works within each spectroscopy type. Before that, in Section~2, we will explain what experimental devices are available. Then, in Section~3~and~4, we will describe how the different types of spectroscopy have been realized with these devices and recount studies. After that, in Section~5, we will summarize the most commonly used theoretical methods.
	
	Some recent reviews lie closely to the topic at hand, and deserves special mention. Rijs and Oomens' 2015 review chapter \cite{rijs2014ir} of gas-phase IR spectroscopy overlaps with this work, although some experimental techniques have emerged since then. Gloaguen and Mons' chapter \cite{gloaguen2015isolated} in the same book focus on a particular technique for IR spectroscopy of neutral gas-phase peptides, which will also be covered in this work. They have also co-authored a longer review \cite{gloaguen2020neutral} on more general spectroscopy of gas-phase neutral peptides. Finally Rijs \emph{et al.} reviews far-IR ($<\SI{800}{\per\centi\meter}$) spectroscopy of neutral peptides in the gas phase. \cite{bakels2020gas}

	\begin{figure*}
		\centering
		\includegraphics[width=\linewidth]{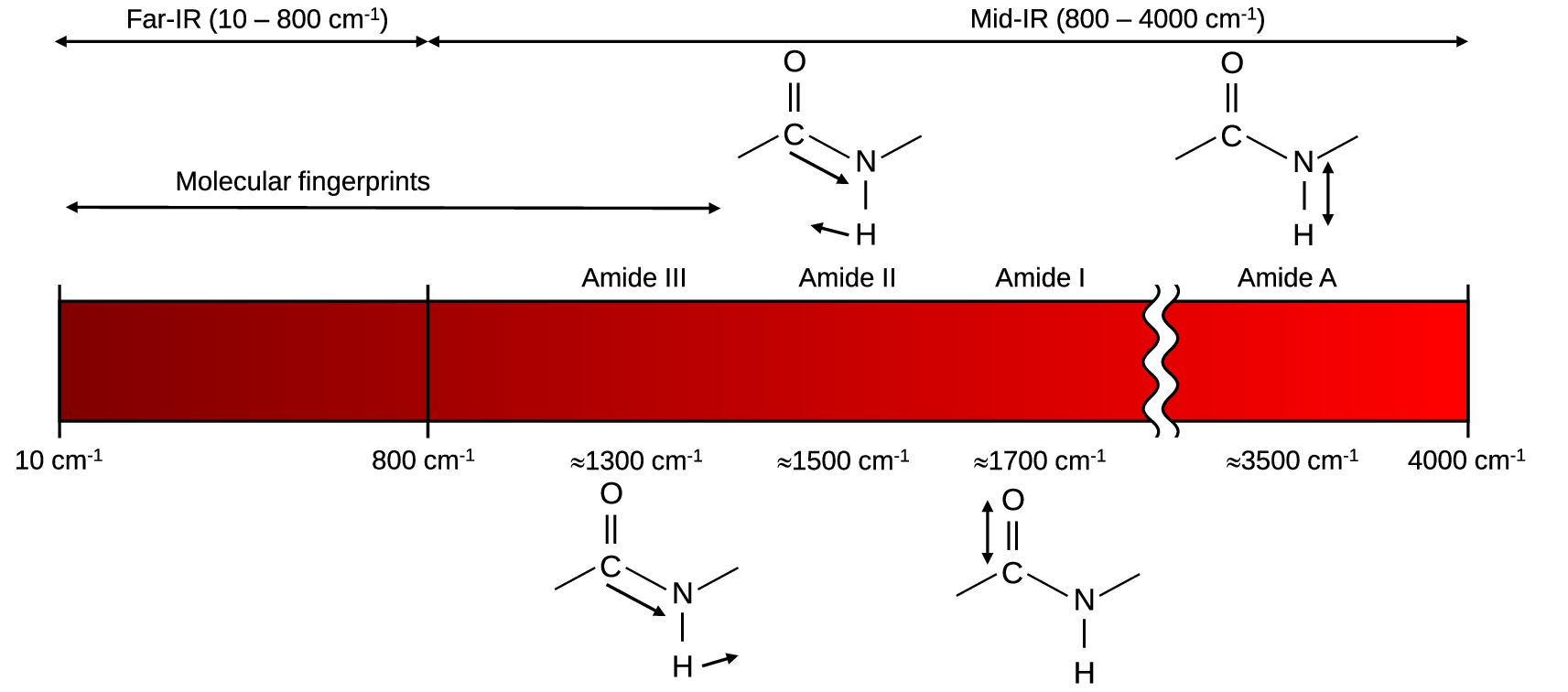}
		\caption{\label{fig:IR}%
			Frequency ranges of common peptide signatures. All peptides contain amides, which have characteristic modes, notably amide A, I, II, and III. More delocalized vibrational modes have larger effective mass and thus resonate at lower frequencies. Vibrational modes that encompass large parts of the molecule are called \emph{molecular fingerprints} because they are collectively molecule-specific.}
	\end{figure*}

	\section{Experimental Devices}
	At least three functions are required in a gas-phase action spectroscopy: First, the molecules must be delivered to the gas phase. Second, the molecules intersect with one or more laser beams. Third and last, some product (such as amount of ions) is counted to infer the absorption rate. The following subsections reflect these three functions, and explain what devices can be used to fulfill each.

	\subsection{Gas-Phase Delivery}
	Depending on the desired charge state of the gas-phase molecule, two techniques are commonly used. Electrospray ionization for charged, and laser desorption for neutral.
	
	\begin{figure*}
		\centering
		\includegraphics[width=0.48\linewidth]{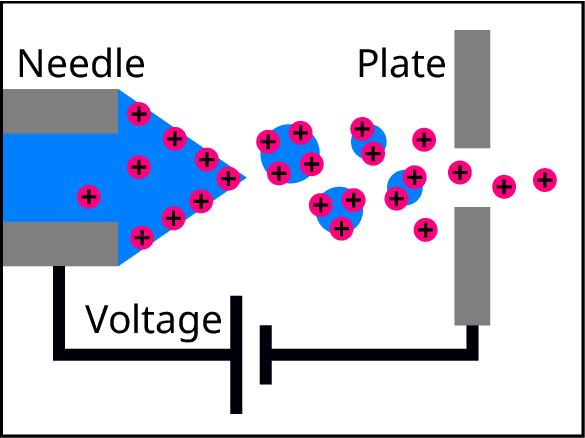}%
		\hfill%
		\includegraphics[width=0.48\linewidth]{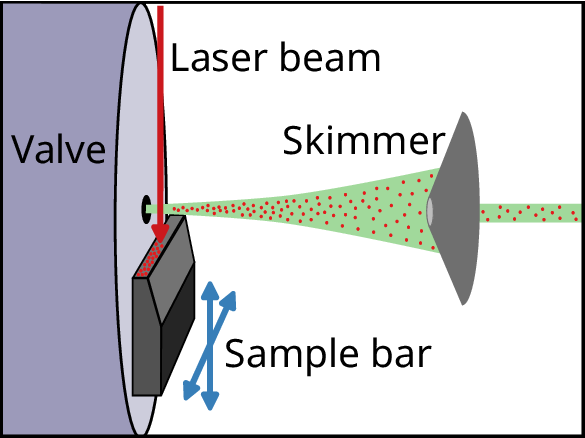}
		\caption{\label{fig:delivery}%
			Techniques for gas-phase delivery of biomolecules. Left: electrospray ionization produces positive ions. Right: laser desorption evaporates molecules which are then seeded into a molecular beam. Adapted from Ref. \cite{ferrari2024infrared} with permission from the PCCP Owner Societies.}
	\end{figure*}
	
	\subsubsection{Electrospray Ionization}
	Electrospray ionization (ESI) delivers molecules to the gas phase in a charged state. \cite{wilm2011principles} To create protonated ions, the molecules are dissolved in a sufficiently acidic solution. The solution flows through a needle held at a positive electric potential (Fig.~\ref{fig:delivery}), and the electrostatic repulsion causes ejection of positively charged droplets, which evaporate and fission into single molecules or small clusters. Deprotonation can be achieved by reversing polarity and using a basic solution.
	
	\subsubsection{Laser Desorption}
	Matrix-assisted laser desorption/ionization (MALDI, or just LD) \cite{karas1985influence} is a technique for delivering molecules to the gas phase.  In MALDI, a short laser pulse is used to locally heat a mixture of the desired molecules and a matrix agent such as carbon powder, added for the purpose of aiding absorption. The added heat lets molecules enter the gas-phase. Because the heating is very brief, the molecules do not break. The absolute majority of molecules remain in a neutral state when delivered to the gas phase. \cite{lu2015ion}
	
	MALDI is often combined with supersonic jet expansion in order to produce a molecular beam, see Fig.~\ref{fig:delivery}. A proportional solenoid valve can be used to emit a short burst of buffer gas which expands in the vacuum chamber. The buffer gas mixes with the desorbed molecules and cools as it expands, creating a conic jet of molecules. A skimmer selects the cold and central part of the jet, which then continues as a molecular beam.

	\subsection{IR Beam Sources}
	As mentioned earlier, due to the minute number of molecules in the interaction region, conventional IR absorption spectroscopy is not applicable. Instead, action spectroscopy is necessary, which places the following requirements on appropriate light sources: relatively narrow linewidth, wavelength tunability, and high intensity. In the field of IR spectroscopy of gas-phase peptides, there are two common approaches to generating IR light that meet these requirements.  They involve making use of table-top lasers and particle-acceleration-based devices such as free electron lasers (FELs). The requirement on high intensity is partially mitigated for the IR action spectroscopy schemes when a single or few IR photons are required. In what follows, these two approaches to generating IR light are discussed in more detail.
	
	\subsubsection{Table-Top Lasers}
	Pulsed nanosecond IR table-top lasers are usually used for IR spectroscopy of gas-phase peptides. They typically comprise three stages, see Refs. \cite{gerhards2004high,bosenberg1993broadly}. The first stage is a powerful Nd:YAG laser emitting $\SI{1064}{nm}$ near-IR light. The linewidth of the emitted light can be as narrow as $\SI{0.01}{\per\centi\meter}$ when the laser is equipped with an injection seeder. The near-IR light is frequency-doubled and serves as a pump for the second stage, the purpose of which is to enable wavelength tunability, that is achieved by employing either an optical parametric oscillator (OPO) or a dye laser. The outcome of the wavelength-tunable laser is directed toward a series of nonlinear IR optical crystals (the third stage) where it overlaps with the residual fundamental light of the Nd:YAG laser. Depending on whether OPO or dye laser is used, the third stage is designed differently. In the case of the OPO laser, the OPO idler light ($1.3$ -- $\SI{2.2}{\micro\meter}$) undergoes optical parametric amplification (OPA). In its turn, the idler light generated by the OPA process ($2.2$ -- $\SI{5.0}{\micro\meter}$) covers the amide A (see Fig. \ref{fig:IR}), NH, and OH stretching bands and, thus, is suitable for spectroscopy of peptides. When the dye laser is used, the third stage contains a difference frequency mixing non-linear crystal to convert the dye laser output into mid-IR light ($1.4$ -- $\SI{4.2}{\micro\meter}$), which is subsequently amplified in an OPA section. Commonly used IR crystals are LiNBO$_3$, KTiOPO$_4$ and KTiOAsO$_4$. The advantage of LiNBO$_3$ is the relatively high non-linear optical coefficient whereas the disadvantages are moisture sensitivity and the spectral gap at around $\SI{2.8}{\micro\meter}$. The latter disadvantage can partially be mitigated by using MgO-doped LiNBO$_3$ crystals. The spectral range can be extended over $\SI{5}{\micro\meter}$ up to $\SI{16}{\micro\meter}$ by difference-frequency mixing of the signal and idler lights of the OPA amplifier with an AgGaSe$_2$ crystal, though the generated thus far-IR light is rather weak to be implemented for action spectroscopy schemes relying on multiple IR photon absorptions. 
	
	\subsubsection{IR Free-Electron Lasers}
	\begin{figure*}
		\centering
		\includegraphics[width=.72\linewidth]{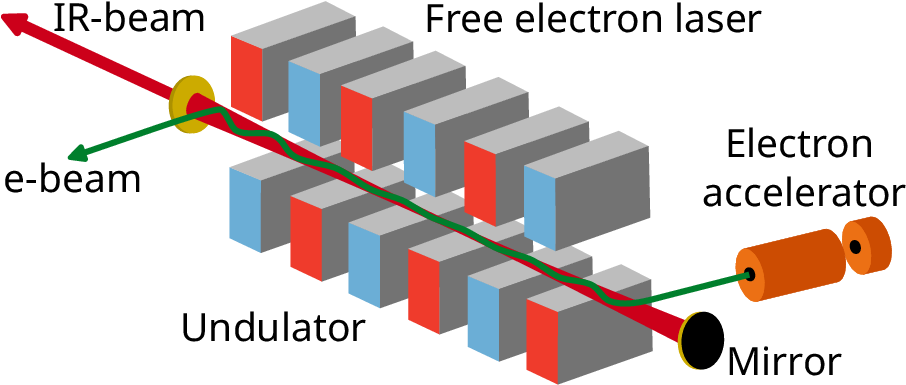}
		\caption{\label{fig:FEL}%
			Schematic overview of an FEL oscillator. Electron bunches are made to oscillate using variable external magnetic fields, and thus emit radiation. Adapted from Ref. \cite{ferrari2024infrared} with permission from the PCCP Owner Societies.
		}
	\end{figure*}
	FELs are large-scale facilities where light is generated by relativistic electron bunches created by radio-frequency linear accelerators (RF LINACs). All FELs comprise undulators which are devices with periodic magnetic fields, see Fig.~\ref{fig:FEL}. Upon traversing the undulator, the electron bunches perform a wiggling motion resulting in light emission. In contrast to conventional lasers, in which the wavelength of the emitted light is determined by energy levels of the active medium, in FELs there are no restrictions on generated wavelength, as the FEL wavelength is determined by undulator parameters (period and magnetic field strength) and electron energy. FELs are designed in two configurations depending on whether it is a single- or multi-pass device, that is, FEL amplifier or oscillator, respectively. FEL oscillators (Fig. \ref{fig:FEL}) comprise optical cavities, where the light generated by an electron bunch is reflected by mirrors to the undulator entrance to interact with a fresh electron bunch and thus amplified. This repeating process continues until the light energy is saturated. IR FELs implemented for IR spectroscopy of gas-phase peptides, such as FELIX \cite{felix}, act as oscillators and are driven by electron micro-bunches at the repetition rate of GHz. Micro-bunches are grouped into $\si{\micro\second}$-long macro-bunches which are generated at the rate of $10$ -- $\SI{20}{\hertz}$. To a first approximation, the FEL linewidth is inversely proportional to the number of undulator periods and is approximately a few \%. However, due to the short-pulse effects, the linewidth of below $\SI{1}{\percent}$ can be achieved by tuning the laser cavity at the expense of FEL gain \cite{bakker1994short}. IR FELs are typically designed to generate IR light in the spectral range below $\SI{2000}{\per\centi\meter}$, which is difficult to cover with IR table-top lasers. FELIX has recently undergone an upgrade and its wavelength range has been extended to $\SI{2.6}{\micro\meter}$ ($\SI{3850}{\per\centi\meter}$).\cite{claessenreport} This recent wavelength extension has opened the door to amide A multiple-photon spectroscopy of chromophore-free gas-phase peptides. Comparing IR table-top lasers and IR FELs, the former are characterized by substantially narrower linewidth, although the latter generate more powerful IR beams and even cover far-IR. 
	
	Because FELs are large and thus expensive, only a few exist in Europe. Notably, Radboud University in Nijmegen has FELIX (Free Electron Lasers for Infrared eXperiments)\cite{felix}, Université Paris Sud in Orsay has CLIO (Centre Laser Infrarouge d'Orsay)\cite{clio}, Fritz-Haber-Institut in Berlin has FHI-FEL (Fritz-Haber-Institut Free Electron Laser)\cite{fhi}. The former two are user facilities, meaning that they regularly let external researchers use the facilities.
	
	\subsection{Mass Spectrometers}
	This report concentrates on action spectroscopy approaches in which IR spectra are measured by monitoring ion production yields versus scanned IR wavelengths. By additionally analyzing masses of ions, mass-specific IR spectra are obtained eliminating thus contributions from spurious ions. The two commonly used methods, conjugated with IR action spectroscopy, are time-of-flight (TOF) and Fourier-transform ion cyclotron resonance (FT-ICR) mass spectrometry. In the following sections, the basic principles behind these two methods are described.
	
	\begin{figure*}
		\centering
		\includegraphics[width=0.60\linewidth]{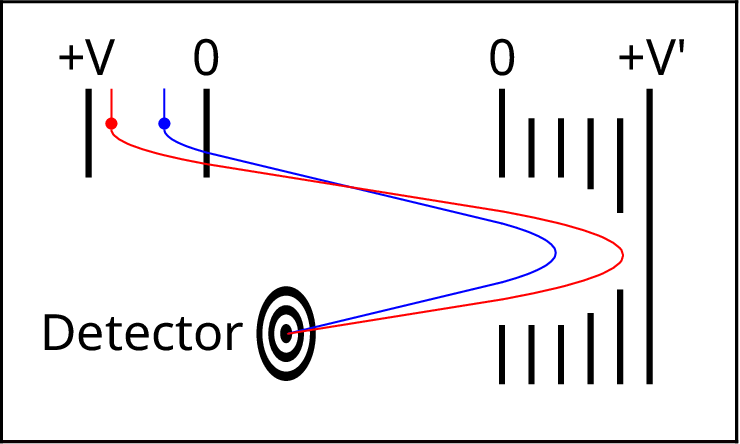}%
		\hfill%
		\includegraphics[width=0.36\linewidth]{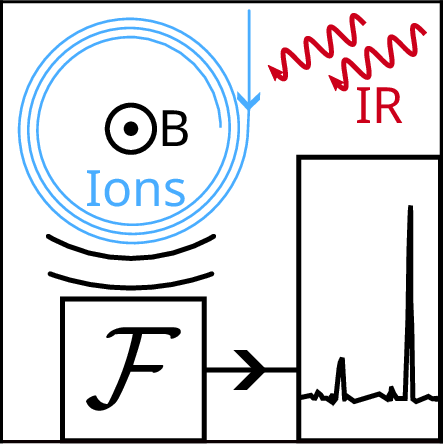}
		\caption{\label{fig:fticr}%
			Schematic presentations of TOF and FT-ICR mass spectrometry methods. Left (TOF): molecules are ionized in the acceleration region of a time-of-flight detector. The reflectron allows a longer drift time by creating another point where the flight time is independent of ionization position. Right (FT-ICR): ions are trapped by a strong magnetic field, and the cyclotron radiation is Fourier transformed into a mass spectrum.
		}
	\end{figure*}
	
	\subsubsection{Time-of-Flight}
	In a TOF mass spectrometer, ions are mass-analyzed based on the time $t$ it takes to traverse a field-free drift region (flight tube) following an interaction region. As these type of mass spectometers require a reference for time measurements, they either operate in a pulsed mode or are used with short-pulse ionizing lasers. An electric field inside the interaction region accelerates ions to nearly identical energies and, as a result, heavier ions travel slower across the flight tube compared to their lighter counterparts. Under these conditions, the TOF $t$ relates to the mass-to-charge ratio of the ion $m/z$ like 
	\begin{equation}
		t = \frac{L}{\sqrt{2eU}}\sqrt{\frac{m}{z}},
	\end{equation}
	where $L$ is the length of the flight tube and $U$ is the acceleration voltage inside the interaction region. The equation implies that provided the flight tube is sufficiently long, different ions can be differentiated according to their TOF. The mass resolution $\Delta m$ relates to the time resolution $\Delta t$ like
	\begin{equation}
		\frac{m}{\Delta m} = \frac{t}{2\Delta t},
	\end{equation}
	and can therefore be further improved by diminishing $\Delta t$. Among the factors influencing $\Delta t$ are detector response time and initial spatial distribution of ions. Due to their relatively fast response time (nanosecond timescale), microchannel plate (MCP) detectors, with two or three plates, are usually implemented in TOF mass spectrometers.  Wiley and McLaren proposed a spectrometer design with compensation for ion spatial distribution. \cite{wiley1955time‐of‐flight} The Wiley--McLaren type mass spectrometer comprises three regions: ionization, acceleration, and field-free drift regions, and acts as an ion focusing lens by adjusting electrical fields in the ionization and acceleration regions. Typically, a mass resolution better than 100 is achieved with Wiley--McLaren mass spectrometers.  An additional order of magnitude in resolution can be obtained by implementing a reflectron-type mass spectrometer. \cite{mamyrin1973mass} This type of mass spectrometers contains an electric reflector placed at the end of the flight tube that acts as an ion mirror thus reversing the velocities of ions, see Fig.~\ref{fig:fticr}. Apart from increasing the length of the flight pass, it further enables diminishing the longitudinal energy spread of ions. 
	
	\subsubsection{Fourier-Transform Ion Cyclotron Resonance}
	When a charged particle is placed in a constant uniform magnetic field $B$, it performs cyclotron motion with the angular frequency 
	\begin{equation}
		\omega = \frac{eB}{m/z}
	\end{equation}
	Fourier-transform ion cyclotron resonance (FT-ICR) mass spectrometers make use of the ion cyclotron resonance phenomenon, in which an ion's mass-to-charge ratio is inferred from its cyclotron frequency. A high magnetic field in the spectrometer is created by a super-conducting magnet. An essential component of an FT-ICR spectrometer is an ICR cell containing, in addition to the magnetic field, an electric field to confine and store the analyzed ions. The cell casing is composed of electrodes, a part of which is used to supply radio-frequency (RF) field and another part is to mirror the charges of the circulating ions.  Being excited by the RF field, the ions increase their cyclotron motion orbits thus getting closer to the detection electrodes. The time-dependent signal measured by the detection electrodes is Fourier-transformed into the frequency-domain spectrum which is then converted to the corresponding $m/z$ spectrum. The main advantage of the FT-ICR mass analyzers is a superb mass accuracy and resolution, which is at least two orders of magnitudes higher than reflectron-type mass analyzers. However, the former are significantly more costly compared to the latter. For further details on FT-ICR mass spectrometry, we refer the reader to Refs. \cite{nikolaev2016fourier,gosseterard2023hyphenation,ruan2023molecular,marshall2015forty}.

	\section{Spectroscopy of Charged Species}
	
	Charged matter responds to electromagnetic fields, enabling manipulation such as trapping and acceleration. Spectroscopy experiments on ions typically use an ESI source to produce ions which are held in a trap during exposure to light. The cyclotron motion in a magnetic trap allow for FT-ICR to measure the mass-to-charge ratio with high accuracy.
	
	\subsection{IRMPD Spectroscopy}
\begin{figure*}
		\centering
		\includegraphics[width=\linewidth]{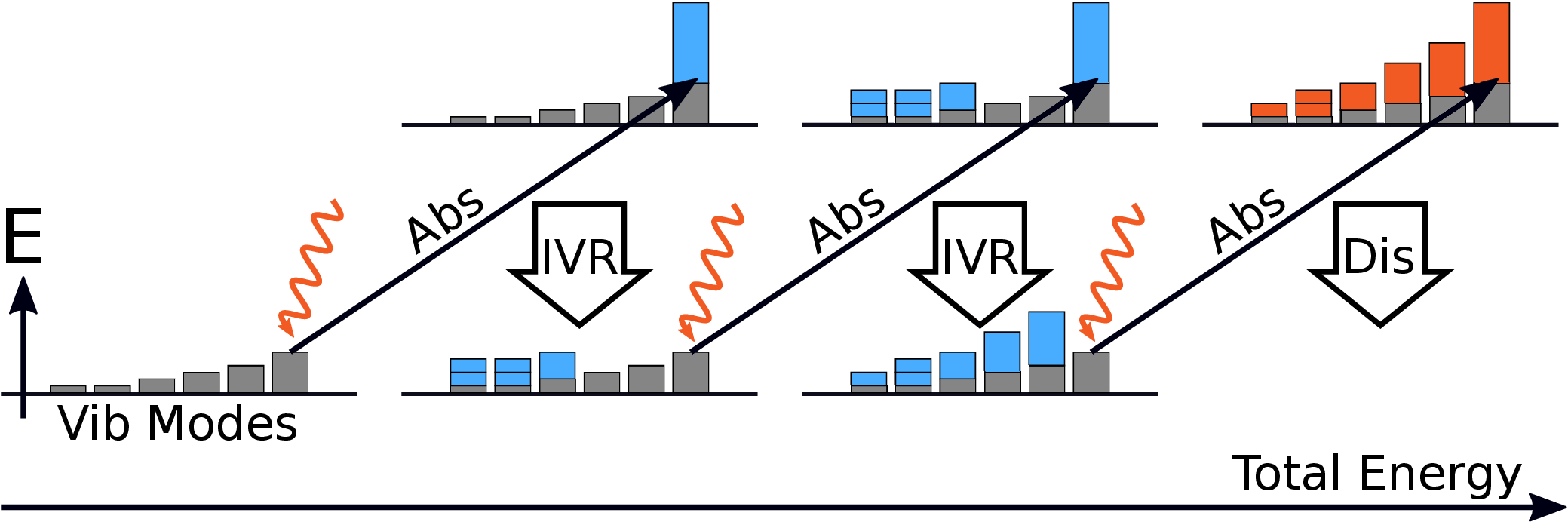}
		\caption{\label{fig:IRMPD}%
			Principle of IRMPD. After a resonant absorption, the frequency shifts due to anharmonic effects. The molecule must undergo intramolecular vibrational energy redistribution (IVR) before another absorption can occur. After a number of IR absorption-IVR cycles, the absorbed energy is great enough to break a bond, and the molecule dissociates.
		}
	\end{figure*}

	Molecules resonantly irradiated with a sufficiently strong IR laser beam dissociate in a process known as IR multiple photon dissociation (IRMPD). Inbetween photon absorptions the molecules must redistribute the vibrational energy from the absorbing mode (see Fig.~\ref{fig:IRMPD}), which takes approximately $\SI{100}{fs}$\cite{lindon2017encyclopedia}. Despite being a multi-photon process, IRMPD yield scales close to linearly with beam power above a certain threshold, because it is limited by the first absorption. \cite{berden2019automatic}
	
	When IRMPD occurs in a trapped ion, the neutral fragments escape the trap and the ion mass effectively decreases. The absorption $\sigma$ can be calculated from the parent ion signal $P$ and summed fragment ion signal $F$ as $\sigma\propto\ln(1+F/P)$. \cite{berden2019automatic}.
	The basic IRMPD experiment on ions consists of an ESI source, an ion trap irradiated by a beam source, and a mass spectrometer. For IR studies with wavenumber less than $\SI{2000}{\per\centi\meter}$ the beam source must be an FEL, such as FELIX, CLIO, and FHI-FEL.
	
	Using FELIX, the group of Oomens and others have performed IRMPD spectroscopy to study protonated \cite{batoon2018conformations,batoon2019characterization} and deprotonated \cite{grzetic2013effect,martinsomer2018unimolecular} short peptides. The purpose is usually to understand the relation between spectrum and structure. Clusters of peptides and metal cations have also been studied \cite{dunbar2013metal,walker2024structural,dunbar2018transition}, motivated by the fact that at least 25\% of proteins bind metal ions\cite{bowman2016metalloprotein}.

	The ability of IRMPD spectroscopy to infer molecular structure can also be used to reveal reaction sites. Such experiments have been performed at FELIX \cite{kempkes2016deamidation,jiang2023oxidation} and CLIO \cite{corinti2023irmpd}.
	Using CLIO, both linear \cite{guan2021gas} and cyclic \cite{hernandez2015rearrangement,alata2017does,barbudebus2022how} peptides have been studied with IRMPD.

	IRMPD action spectroscopy can be combined with ion mobility to give even more information about the structure. \cite{warnke2015analyzing} Ion mobility implies that the ions are dragged by a weak homogeneous electric field through a drift tube containing a low-pressure buffer gas. Ions with a large collision cross section take longer to pass through the tube. Thus extended conformers will be time-separated from folded ones.

\begin{figure}
		\centering
		\includegraphics[width=\linewidth]{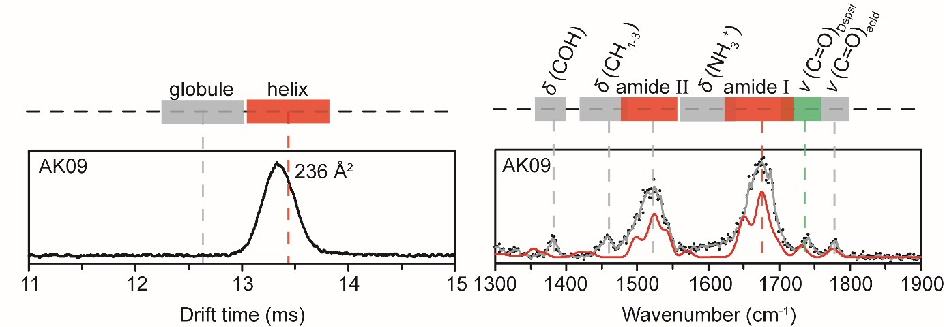}
		\caption{\label{fig:drift_time}%
			Combined ion mobility and IRMPD spectroscopy of ester-substituted Ac-Ala$_{10}$-Lys-H$^+$. Left: The measured drift time spectrum rules out globule-like conformers. Right: The IRMPD spectrum (gray) compared to the quantum chemical prediction (red).
			Adapted from Ref. \cite{hoffmann2016assessing} with permission from the PCCP Owner Societies.
		}
	\end{figure}

	Using FHI-FEL, combined ion mobility and IRMPD spectroscopy studies of polyalanine peptides \cite{schubert2015exploring,hoffmann2016assessing} have been performed. 
	Fig.~\ref{fig:drift_time} shows the result of such an experiment. \cite{hoffmann2016assessing} The measured drift time of ester-substituted Ac-Ala$_{10}$-Lys-H$^+$ rules out all globule-like conformers, and the IRMPD spectrum gives additional information.
	
	\subsection{Messenger-Tagging IR Spectroscopy}
	Another way to realize action spectroscopy is to start with an adduct of the ion of interest and a weakly bound tag molecule, usually N$_2$ or H$_2$. Upon absorption of a single IR photon the tag will be separated from the ion, causing its mass to decrease. The absorption can be calculated from the ions signals as in IRMPD, using the adduct as the parent ion.
	
	The main advantage of messenger-tagging IR spectroscopy is a diminished demand on high intensity for implemented IR light sources. However, a known downside to this technique is that  the tag molecule alters the structure of the ion. The magnitude of this effect is estimated in a study of protonated tryptophan. \cite{spieler2018vibrational} By comparing spectra of clusters TrpH$^+$(N$_2$)$_n$ for $n=1$ -- $5$, it is seen that adding a single N$_2$ tag induces frequency shifts in some bands related to the binding site.
	
	To our knowledge, only a few peptides have been investigated with messenger-tagging.	The dipeptides AlaTyrH$^+$ and TyrAlaH$^+$ have been studied with water as the tag \cite{saparbaev2022identification}, and GlyGlyH$^+$ with N$_2$ \cite{chen2021time}. Also, the eight tripeptides [Gly,Ala]$_3$H$^+$ have been studied using D$_2$ as the tag \cite{sherman2022conformational}.
	
	Messenger-tagging IR spectroscopy in combination with ion mobility has been developed by the group of Rizzo and applied to glycans (polymers of saccharides). \cite{khanal2017glycosaminoglycan,warnke2019combining,bansal2020using,warnke2021high} The setup allows for glycan ions to be separated by collisional cross-section and selected by mass before being tagged and irradiated. A similar setup in Amsterdam has recently been developed and tested on peptides by the group of Rijs. \cite{bakels2024probing}
	
	\section{Spectroscopy of Neutral Species}
	
	Because neutral matter cannot be manipulated by electromagnetic fields, spectroscopy of neutral molecules must involve ionization. Thus, IR spectroscopic methods can be classified by how the molecule is ionized. We will in the following subsections consider resonant two-photon and one-photon ionization.
	
	\subsection{IR--UV Ion-Dip Spectroscopy}
\begin{figure*}
		\centering
		\includegraphics[width=.3\linewidth]{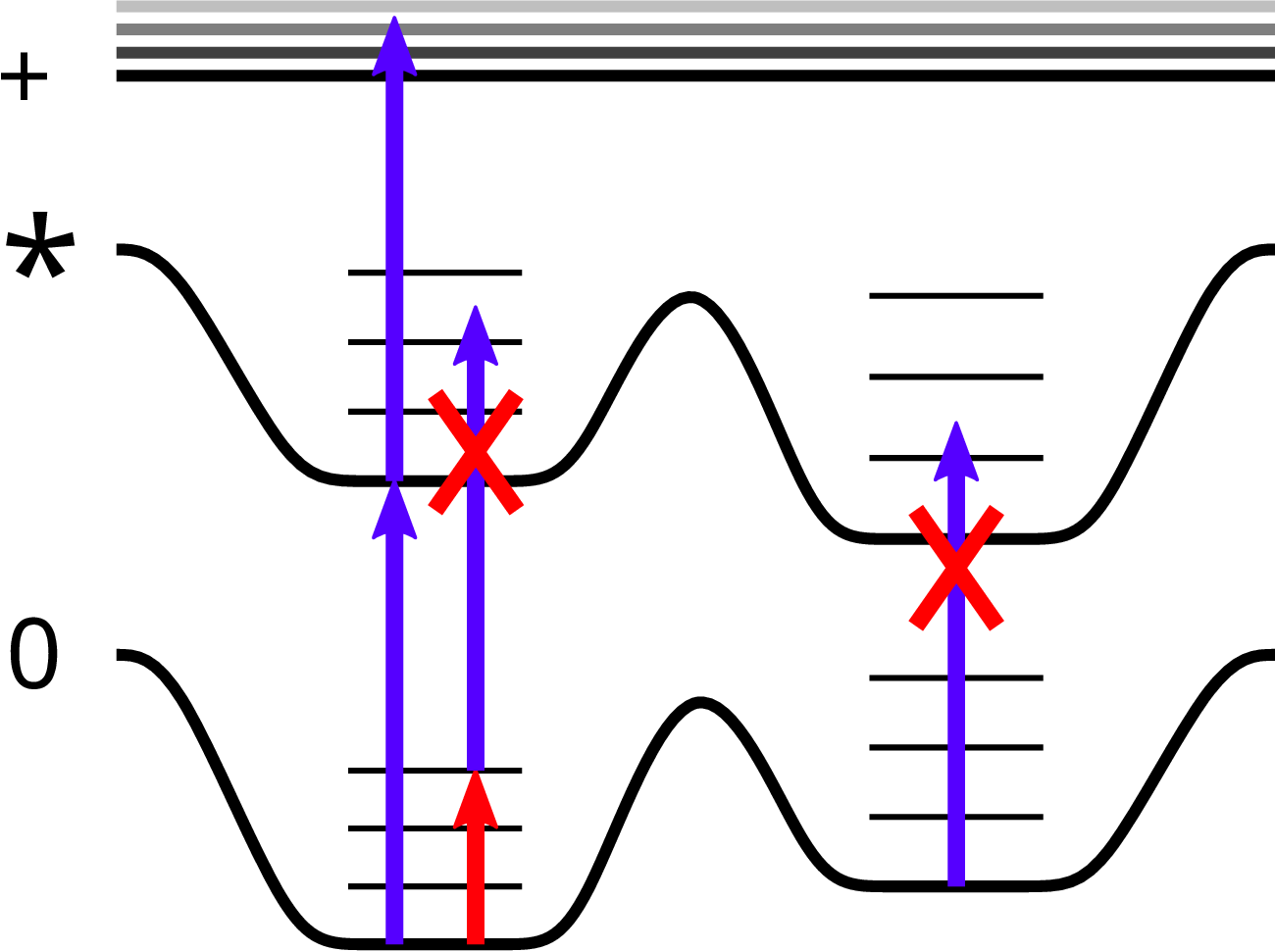}%
		\hfill%
		\includegraphics[width=.3\linewidth]{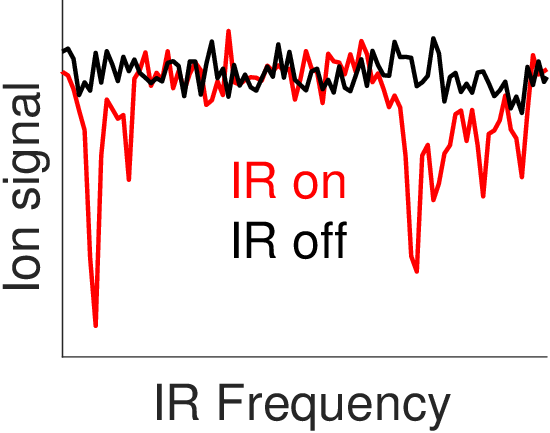}%
		\hfill%
		\includegraphics[width=.3\linewidth]{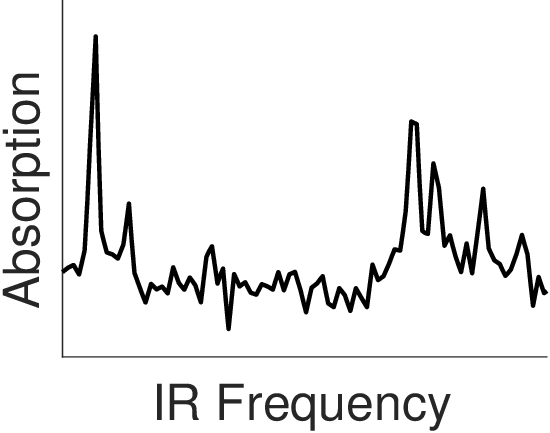}
		\caption{\label{fig:IRUV}%
			Principle of conformer-specific IR--UV ion-dip spectroscopy. Two UV photons can ionize the molecule resonantly, but only from a specific conformer in the ground state. IR absorption depopulates the ground state, and thus decreases the ion signal. By measuring the ion signal with and without IR fluence, the absorption can be estimated.
		}
	\end{figure*}

	In IR--UV ion-dip spectroscopy (Fig.~\ref{fig:IRUV}), also known as resonant ion-dip IR (RIDIR) spectroscopy, there are two types of photons: IR photons that excite vibrations, and UV photons that resonantly excite and subsequently ionize ground state molecules. Without IR light, resonant two-photon ionization (R2PI) occurs, more generally called resonance-enhanced multi-photon ionization (REMPI), giving an ion signal. Because R2PI is resonant and the intermediate state has a relatively long lifetime, a specific conformer can selectively be ionized. When the IR light preceding the UV light is absorbed, the R2PI process is inhibited in vibrationally excited molecules, causing a dip in the ion signal. By combining the ion signal $P$ with and without IR, the absorption $\sigma$ can be calculated as $\sigma\propto\ln(P_\mathrm{without IR}/P_\mathrm{with IR})$. Thus IR--UV ion-dip spectroscopy gives a conformer-specific single-photon IR spectrum.
	
	For R2PI to be possible, the molecule must contain a UV-absorbing chromophore. The most popular implementation in peptides is the residue phenylalanine, whose side chain contains a benzene ring that absorbs UV at approximately $\SI{267}{nm}=\SI{37500}{\per\centi\meter}$.\cite{alauddin2014intra} Other options with similar wavelengths are the N-terminal cap benzyloxycarbonyl,\cite{kumar2019observation} the C-terminal cap benzyl,\cite{kusaka2013role} or the residues tyrosine and tryptophan. This wavelength can be reached by a Coumarin 153 dye laser combined with a frequency doubling crystal.
	
\begin{figure*}
		\centering
		\includegraphics[width=\linewidth]{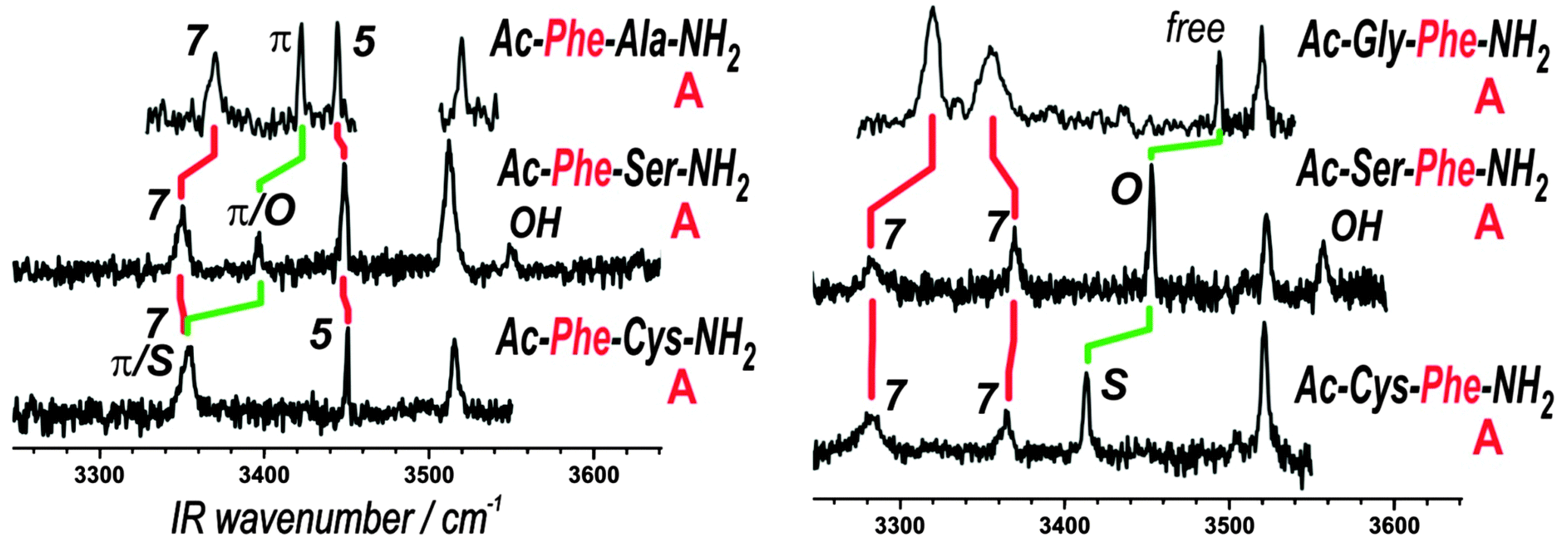}
		\caption{\label{fig:mons_adapted}%
			IR--UV ion-dip spectra of capped dipeptides. The phenylalanine residue contains a chromophore which permits resonant UV absorption. By varying the other residue, effects of the side chain is seen.	Adapted from Ref.~\cite{alauddin2014intra} with permission from the PCCP Owner Societies.
		}
	\end{figure*}

	The group of Mons has extensively studied neutral capped short peptides containing phenylalanine by performing IR--UV ion-dip spectroscopy in the $3200$ -- $\SI{3600}{\per\centi\meter}$ frequency range, which covers stretching modes and can be realized with table-top lasers. \cite{gloaguen2020neutral} Examples include the monopeptide Ac-Phe-NH$_2$ \cite{sohn2016local}; the dipeptides Ac-Phe-Xxx-NH$_2$ and Ac-Xxx-Phe-NH$_2$ for Xxx = Ala, Asn, Cys, Ser, His, and Phe \cite{loquais2015secondary,alauddin2014intra,sohn2016local,habka2018turn}; the tripeptides Ac-(Ala)$_2$-Phe-NH$_2$, Ac-Ala-Phe-Ala-NH$_2$, and Ac-Phe-(Ala)$_2$-NH$_2$ \cite{chin2005gas}; and the tetrapeptides Ac-(Ala)$_3$-NH$_2$ \cite{plowright2011compact}. By comparing the spectra of similar peptides, it is possible to understand the effect of individual residues, see Fig.~\ref{fig:mons_adapted}.
	
	Although not natural, peptides containing $\beta$- and $\gamma$-amino acids have been studied by the group of Zwier. \cite{walsh2013cyclic,gord2014mimicking,fischer2022conformer}.
	Other neutral peptides studied with IR--UV ion-dip spectroscopy in the stretching range include
	Z-Gly-Pro-OH ,\cite{kumar2019observation}
	Z-(Aib)$_n$-OMe ,\cite{gord2016conformation}
	Ser-Ile-Val-Ser-Phe-NH$_2$ ,\cite{ishiuchi2016gas}
	Boc-Gly-(\textsc{d}-Pro)-NHBn-OMe, and
	Boc-(\textsc{d}-Pro)-Gly-NHBn-OMe \cite{kumar2022sequence}
	.
	
	Using FELIX, the group of Rijs have performed IR--UV ion-dip spectroscopy to study neutral peptide dimers in the far-IR range. Examples include Ac-Phe-Xxx-NH$_2$ for Xxx = Gly, Ala \cite{jaeqx2014gas}, Pro\cite{mahe2015can}, Cys, Ser, Val \cite{mahe2017mapping}; the uncapped dimers PhePhe and PhgPhg \cite{stroganova2022structural,bakels2021probing}; and clusters of Ac-Phe-OMe \cite{galimberti2019conformational}.
	These experiments were accompanied by Born--Oppenheimer molecular dynamics generation, a theoretical method described later in this chapter.

	\subsection{IRMPD--VUV Spectroscopy}
\begin{figure*}
		\centering
		\includegraphics[width=.9\linewidth]{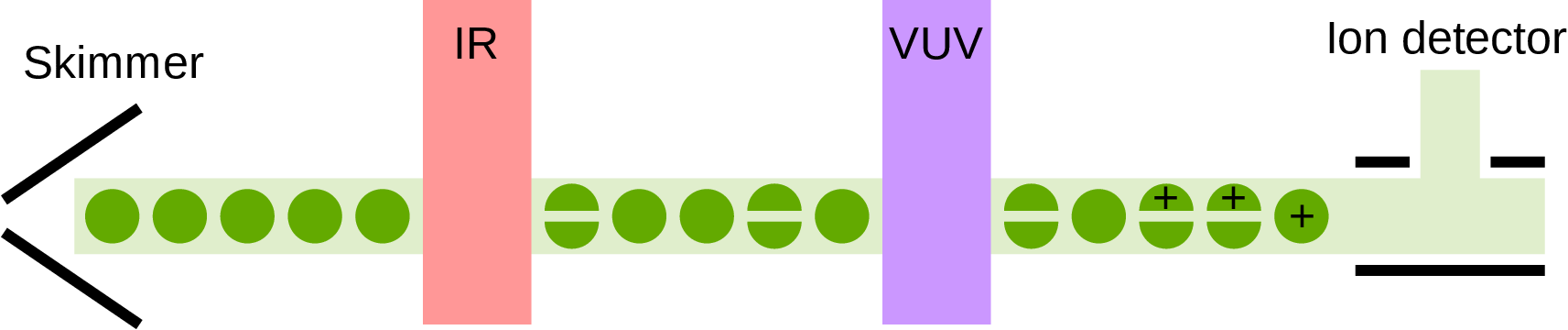}
		\caption{\label{fig:IRMPD_VUV}%
			Schematic overview of IRMPD--VUV experiment. A neutral molecular beam is first partially fragmented by IRMPD and then partially ionized and fragmented by VUV photon absorption. By measuring the ion signal with and without IR light, the absorption can be estimated.
		}
	\end{figure*}

	IRMPD--VUV spectroscopy is a relatively new technique \cite{yatsyna2016infrared} which allows study of neutral molecules without chromophore, see Fig.~\ref{fig:IRMPD_VUV}. As the name alludes to, neutral molecules in a beam are irradiated with two light pulses in sequence: first with intense IR light, and then with vacuum ultraviolet (VUV) light. The IR pulse causes some molecules to fragment via IRMPD, and the VUV pulse ionizes some molecules and fragments via single-photon absorption. By measuring the parent ion signal $P$ and summed fragment ion signal $F$ with and without IR light, the absorption $\sigma$ can be estimated to be
	\begin{equation}
		\sigma \propto \ln\left(1+\frac{F_\mathrm{with IR}}{P_\mathrm{with IR}}\right) - \ln\left(1+\frac{F_\mathrm{without IR}}{P_\mathrm{without IR}}\right).
	\end{equation}
	The reasoning behind this formula is given in Ref.\cite{andersson2023indication}.
	
	Only a few peptides have been studied with the IRMPD--VUV method. The groups of Zhaunerchyk and Rijs collaborated in studies of Gly$_2$ \cite{yatsyna2019conformational}, Ala$_2$ \cite{yatsyna2019competition}, and Ala$_5$ \cite{andersson2023indication}. These experiments used a $\SI{118}{nm} = \SI{84600}{\per\centi\meter}$ VUV laser beam produced by a Xe--Ar cell in combination with FELIX. The results showed that the dipeptides assumed extended conformations, but the pentapeptide preferred a $3_{10}$-helix-like structure.
	
	A drawback of the IRMPD--VUV method is the lack of conformer selectivity, which leads to spectral congestion in large molecules such as Ala$_5$.
	
	\section{Theoretical Methods}
	
	Three questions always arise in an IR spectroscopic study of a peptide: What are the plausible conformers, what are their relative abundances, and what are their IR spectra? The first question is particularly difficult to answer for peptides, because their large conformational freedom makes an exhaustive listing of conformers intractable, and requires specialized algorithms with qualified assumptions. The latter two questions can be answered with quantum chemical methods common to all biomolecules, with density functional theory (DFT) being the most popular option.
	
	The workflow for determining the IR spectrum of a given peptide naturally follows these three questions in order. First, some algorithm is used to generate conformations. These are then refined into conformers, which by definition are minima on some potential energy surface (PES). Second, the relative abundances of the conformers are estimated from the highest-quality energy calculations affordable. Third and last, the IR spectra of the most abundant conformers are calculated.
	
	\subsection{Conformational Search Methods}
	
	The first and most difficult task in the theoretical investigation of a peptide is the conformational search. It is difficult because the conformational space, which by definition is generated by dihedral angle rotations around single bonds, is exponentially large. Studying even the simplest peptide Gly-Gly with 6 rotable bonds in increments of $\SI{30}{\degree}$ would require the sampling of $12^6 \approx \SI{3e6}{}$ conformations, a prohibitively large number for all but the most cursory methods. Hence, more sparse sampling strategies of the conformational space must be used.
	
  	\subsubsection{Sparse Uniform Sampling}
	A simple way to generate conformations is to randomly choose each dihedral angle by sampling a distribution on $[-\SI{180}{\degree},\SI{180}{\degree}]$ \cite{yatsyna2019conformational,yatsyna2019competition} or a discrete subset thereof. The distributions can be non-uniform; in \emph{knowledge-based} methods, they depend on the chemical context, applying assumptions such as peptide links almost always being near planar ($\omega = \SI{180}{\degree}$), or hydrocarbon chains preferring staggered (trans or gauche) configurations. {Confab} \cite{oboyle2011confab} is a knowledge-based \cite{ebejer2012freely} free implementation of sparse uniform sampling.
	
	Randomly choosing angles may produce unphysical self-intersecting conformations, especially when applied to peptides. This inefficiency is solved by \emph{distance geometry embedding} methods, which start out by constructing upper and lower bounds on all atom-pair distances based on bond data. A target distance matrix is then randomized within the bounds, and the conformation is manipulated so that its distance matrix is as close as possible to the target. This method and some of its variations are freely available in the software package {RDKit}. \cite{landrum2010rdkit} Notably, knowledge-based post-processing is needed to flatten out aromatic rings and sp$^2$ centers. \cite{riniker2015better}
	
	\subsubsection{Energy-Guided Sampling}
	The search methods mentioned so far do not consider energy, and therefore sample low- and high-energy conformations at equal rate. This is sometimes desirable when one wants to explore an entire conformational space, but has the downside of generating many conformations that scarcely occur at experimental temperature. The alternative is to focus on low-energy conformer by incorporating intermediate energy or force calculations. Molecular force field methods such as MM3 \cite{allinger1989molecular}, are appropriate for this task because they are very fast to run.
	
	Generation of low-energy conformations obeying a Boltzmann distribution can be achieved with the Metropolis--Hastings algorithm \cite{hastings1970monte,lin2017markov}, in which the conformation is stochastically mutated dependent on energies. Mutations can be dihedral angle changes \cite{hung2010de} or exchanges \cite{sakae2014conformational}. In some variants the structure is intermediately optimized. \cite{li1987monte,vasquez1994free} Redundancy in generation can be avoided with \emph{tabu search} rules at the cost of some memory. \cite{grebner2011efficiency, glover1998tabu}
	
	Another way to generate conformers without redundancy is \emph{basin-hopping} as implemented in the software package {Tinker}. \cite{rackers2018tinker} The method starts with a queue containing only the initial conformer. That queue is then iterated through, letting each conformer be kicked along normal modes and optimized to hopefully discover new conformers, which are then appended to the queue. In practice the method does not terminate and an arbitrary conformer limit is used.
	
	\subsubsection{Dynamical Sampling}
	Perhaps the most obvious way to obtain conformers is through physical molecular dynamics (MD) simulations, by periodical sampling and optimization. In theory, a single simulation on an anharmonic surface eventually explores the entire space as time goes to infinity. In practice this is inefficient, and multiple simulations with different initial conditions are run instead. Physical MD with the popular Verlet integrator \cite{hairer2003geometric} requires only force calculations and is implemented in most if not all quantum chemistry packages.
	
	Beyond the physical simulations there are many metadynamical tricks that improve the sampling. \emph{Annealing}, meaning gradually lowering the temperature by re-scaling velocities, permit more dense generation of conformers while still overcoming barriers. In \emph{replica-exchange} MD \cite{sugita1999replica,kotobi2023reconstructing}, multiple simulations run in parallel and randomly exchange temperatures. The {CREST} software \cite{pracht2024crest} avoids redundancy by adding to the MD potential a repelling bias for each conformer already found.

	\subsubsection{Machine Learning Efforts}
	In 2020, the deep neural network AlphaFold2 gained fame as it won a protein structure prediction challenge by a large margin. \cite{jumper2021applying,jumper2021highly,abramson2024accurate} It leverages genetic databases to encode the primary sequence before input, and predicts the 3D structure with a median backbone RMSD of $\SI{2.32}{\angstrom}$. \cite{jumper2021highly} AlphaFold2 performs similarly on peptides down to a residue length of 10. \cite{mcdonald2023benchmarking} The newly-released sequel AlphaFold3 \cite{abramson2024accurate} may perform better, but no such benchmarking has been done yet.
	
	\subsection{Energy and Force Calculations}
	
	There are a variety of quantum chemical models along the axis of simplicity--accuracy. In a conformational search, the model must be simple to permit many evaluations, and typically rely on empirical data. When refining the structure, a more accurate model is typically used to enforce realistic behaviors. Estimating abundances requires an accuracy on the order of $k_\mathrm{B}T$, which is only reached by the most expensive models.
	
	\subsubsection{Molecular Force Fields}
	The simplest models are molecular force fields (FFs), in which the energy is assumed to be an analytical function of molecular distances and angles. As an example, the AMBER model \cite{weiner1981amber} writes the energy as a sum of polynomial functions of bond lengths and angles, cosines of dihedral angles, and the Lennard--Jones and Coulomb potential of interatomic distances. Over the years more complex FFs have been created: CHARMM \cite{brooks1983charmm}, MM3 \cite{allinger1989molecular}, MMFF \cite{halgren1996merck}, GAFF \cite{wang2004development}, and ECEPP \cite{arnautova2006new} to name a few. All these models have coefficients which are fitted to empirical measurements or higher-level calculations of sets of molecules. While FFs aim to give a qualitatively correct PES, their energies correlate poorly with DFT. \cite{kanal2018sobering}
	
	Machine learning can been used to systematically construct more complex FFs, and are typically one or two orders of magnitude slower than simple FFs. \cite{behler2016perspective} \emph{Kernel-based} models \cite{bartok2015gaussian, glielmo2017accurate, chmiela2017machine, sauceda2019molecular} use descriptors such as the distance matrix and the Coulomb potential matrix to encode molecular structure, and then combines all pairs of descriptors into a kernel matrix, which maps a trainable parameter vector to an observable, typically force or energy. \emph{Neural network} models \cite{behler2007generalized,jose2012construction,mardt2018vampnets,dong2024searching} use a feed-forward network to map the chemical environment of each atom or atom-pair to some local quantity.
	
	\subsubsection{Semiempirical Methods}
	Beyond FFs in terms of complexity are the semiempirical methods, which consider wavefunctions and Hamiltonians but make approximations and take empirical parameters. The neglect of diatomic differential overlap family of methods include AM1 \cite{dewar1985development}, RM1 \cite{rocha2006rm1}, and most recently PM7 \cite{stewart2013optimization} which are all freely available in the {MOPAC} software by Stewart.\cite{stewart2024mopac} When calculating heats of formation, PM7 has a typical error of $\SI{8.52}{kcal/mol}$ in general \cite{stewart2024mopac}, and correlates to $20$ -- $\SI{40}{\%}$ with DFT on conformers within the same molecule \cite{kanal2018sobering}.
	
	For the purpose of analyzing large proteins, some methods with linear complexity have been developed. Stewart's {MOZYME} \cite{stewart2009application} makes some time-saving simplifications to the PM6 method. A method similar in spirit but starting from localized molecular orbitals is {LocalSCF}. \cite{anikin2004localscf} In a comparison made by its creators, LocalSCF has better energy accuracy than MOZYME on conformers of hydrated insulin. \cite{anisimov2006validation}
	
	\subsubsection{Higher Level Methods}
	Next level in terms of complexity is density functional theory (DFT), which posits that the system energy is a functional of the electron density. This assumption is made to avoid handling the electron wavefunction, which is far too large.
	The runtime of DFT scales with the system size cubed. \cite{perdew2001jacobs} Generalized gradient approximation (GGA) functionals such as BLYP\cite{becke1988density,lee1988development}, BP86\cite{becke1988density,perdew1986density}, and PBE\cite{perdew1996generalized} are relatively cheap and find use in molecular dynamics simulations. The more expensive hybrid functionals such as B3LYP\cite{becke1993density‐functional}, PBE0\cite{adamo1999reliable}, M06-2X\cite{zhao2008m06}, or $\omega$B97X-D\cite{chai2008long,lin2013long} are now commonly employed for geometry optimization and frequency calculation of peptides. 
	
	M{\o}ller--Plesset methods \cite{cremer2011moller} are sometimes used for single-point calculations. The most common, MP2, scales with the system size to the power of five. Double-hybrid methods such as B2PLYP\cite{grimme2006semiempirical} combine DFT with an MP2-like correction for improved accuracy.
	
	Finally, methods of chemical accuracy like CCSD(T) \cite{purvis1982full,pople1987quadratic} and G4(MP2) \cite{curtiss2007gaussian} are only affordable for small peptides.

	\subsection{Frequency Calculations}
	The standard method for calculating vibrational frequencies is making a harmonic approximation of a hybrid functional. Empirical scaling factors are needed to compensate for the lost anharmonicity. After scaling, the typical mean absolute error is around $\SI{30}{\per\centi\meter}$ depending on the functional. \cite{halls2001harmonic}
	
	A more complex alternative is vibrational perturbation theory \cite{franke2021how}, most often of the second order (VPT2). VPT2 considers up to quartic derivatives of the PES, and is usually more accurate than the harmonic approximation. \cite{yatsyna2016aminophenol,biczysko2010harmonic,xu2024harmonic} While a full VPT2 requires the computation of $\mathcal{O}(N^3)$ coefficients, there are schemes for reducing the cost, enabling application to larger systems. \cite{bloino2023anharmonicity,fuse2024scaling}
	
	One major flaw with VPT2 is that the expression $\omega_i+\omega_j-\omega_k$ (where $\omega_i$ are harmonic vibrational frequencies) appears in some denominators, leading to absurdly large terms whenever $\omega_i+\omega_j\approx\omega_k$, known as Fermi resonances. The removal of such terms can be implemented in many ways. \cite{yang2022effective, bloino2023reliability}
	
	\subsubsection{BOMD Spectrum Generation}
	The methods discussed so far depend on a single geometry, which is ill-fitting for systems such as floppy peptides and hydrated complexes. A proposed method for such systems is Born-Oppenheimer molecular dynamics (BOMD) spectrum generation. BOMD implies that the nuclei are simulated classically subject to Hellman--Feynman forces. The result of BOMD is trajectories exploring the phase space. These trajectories are then processed to yield an IR spectrum. \cite{thomas2013computing,gaigeot2014theoretical}

	Specifically, the BOMD method uses the Fermi golden rule to relate the absorption spectrum to the power spectral density of the centered electric dipole $\tilde{\mu}$:
	\begin{equation}
		\sigma(\omega) \propto 
		\omega^2
		\int_{-\infty}^\infty d\tau\, e^{i\omega\tau}
		\int_{-\infty}^\infty dt\, \tilde{\mu}(t)\cdot \tilde{\mu}(t+\tau).
	\end{equation}
	In practice, the integrals are limited by the simulation duration. In the limit of an infinite simulation duration and zero temperature, the harmonic spectrum is reproduced.
	
	Some method parameters must be chosen by hand. Temperature and initial conditions can be physically motivated \cite{vanoanh2012improving}; but time step, duration, and number of runs are purely computational. The time step should be much smaller than the shortest vibrational period to guarantee fidelity; using a time step too large causes a frequency shift. \cite{andersson2023indication,andersson2024universal} The duration should be at least as long as the inverse of the desired spectral resolution. The number of runs can be increased to improve statistics, but is ultimately limited by computational resources.
	
	The frequency shift due to a large time step is independent of the molecule and predictable. \cite{andersson2023indication,andersson2024universal} In theory, it should be possible to run BOMD with a large time step to decrease runtime, and then reverse the frequency shift. However, the practical validity of such a strategy has not been demonstrated.
	
	While the BOMD method has been successful for some molecules \cite{jaeqx2014gas}, it has a principal flaw: The anharmonic frequency shift scales with simulation temperature. As pointed out in Ref.~\cite{suhm2013femtisecond}, to correctly reproducing the fundamental frequency $\omega_0$ of a mode requires a temperature of $\hbar\omega_0/k_\mathrm{B}$, more than an order of magnitude greater than what is typically used. The BOMD method should therefore not be thought of as an anharmonic method.
	
	\section{Summary}
	
	This review chapter has summarized the current state of IR action spectroscopy of gas-phase peptides. 
	We have given an overview of the necessary experimental devices to investigate proteins with gas-phase IR spectroscopy. Molecules are put into the gas phase by some mechanism such as ESI or MALDI, and trapped if possible. While in the gas phase, the molecules absorb light, causing some actions such as fragmentation or ionization. Some resulting quantity, typically ion count and mass, is then measured by a spectrometer to infer the absorption rate.
	
	Several types of IR action spectroscopy are possible, depending on what information is desired and what instruments are available. IRMPD spectroscopy requires intense IR light to fragment molecules, typically charged but recently also neutral (IRMPD--VUV). Messenger-tagging IR spectroscopy produces single-photon spectra by dissociating ion--tag complexes.	
	IR--UV spectroscopy allows for conformer-specific IR spectra to be obtained. 
	
	We have also listed and briefly explained the most common theoretical methods for predicting the IR spectra. A thorough analysis of a molecule requires a conformational search, followed by an energy ranking of the found conformers, and then frequency calculations of the lowest-energy conformers. The optimal method is different for each task.
	
	In conclusion, gas-phase IR spectroscopy of peptides has proven to be a powerful tool for probing the vibrational spectrum of peptides. When combined with quantum chemical methods, this spectrum can be used to discriminate and confirm molecular structures. Eventually, such confirmed structures will be used to train new semiempirical methods. This synergy between experiment and theory paves the way to a bottom-up understanding of protein mechanics.
	
	\bibliographystyle{unsrt}
	\bibliography{references}
\end{document}